\documentclass[conference]{IEEEtran}
\IEEEoverridecommandlockouts
\usepackage{cite}
\usepackage{amsmath,amssymb,amsfonts}
\usepackage{graphicx}
\usepackage{textcomp}
\usepackage{xcolor}
\usepackage{balance}
\usepackage{algorithmic}
\usepackage[ruled,vlined,linesnumbered]{algorithm2e}
\usepackage[disable]{todonotes}
\usepackage{booktabs}

\newcommand{\IOpub}{\texttt{IOpub}}

\begin{document}
\title{Context-aware Execution Migration Tool for\\ Data Science Jupyter Notebooks on Hybrid Clouds}
\author{\IEEEauthorblockN{Renato L. F. Cunha, Lucas V. Real, Renan Souza, Bruno Silva, Marco A. S. Netto}
\IEEEauthorblockA{IBM Research
}
}

\maketitle

\begin{abstract}

Interactive computing notebooks, such as Jupyter notebooks, have become a
popular tool for developing and improving data-driven models. Such notebooks
tend to be executed either in the user's own machine or in a cloud
environment, having drawbacks and benefits in both approaches. This paper
presents a solution developed as a Jupyter extension that automatically
selects which cells, as well as in which scenarios, such cells should be
migrated to a more suitable platform for execution. We describe how we
reduce the execution state of the notebook to decrease migration time and we
explore the knowledge of user interactivity patterns with the notebook to
determine which blocks of cells should be migrated. Using notebooks from
Earth science (remote sensing), image recognition, and hand written digit
identification (machine learning), our experiments show notebook state
reductions of up to 55$\times$ and migration decisions leading to performance
gains of up to 3.25$\times$ when the user interactivity with the notebook is
taken into consideration.
\end{abstract}


\section{Introduction}

Interactive computing notebooks, such as Jupyter notebooks, have become popular
for developing and tuning data-driven
models~\cite{perkel2018jupyter,shen2014interactive}.  They are related to
\emph{literate programming}~\cite{knuth1984literate} and \emph{literate
computing}~\cite{millman2018developing} paradigms, which aim at helping the
communication of computer programs by keeping code and documentation closely
together, and storing results of a code execution in a document, along with
figures, tables, and free-form text, respectively. They have become a useful
tool for scientists and engineers to share code, prototype models, and analyze
their results. They are also seen as a mechanism to help reproducing
results~\cite{kluyver2016jupyter,paine2020understanding}, increasing
collaboration~\cite{mendez2019toward}, and facilitating usage of complex HPC
applications~\cite{nishimura2017lattice}.

 A Jupyter notebook contains rectangular cells in a web application---a cell can
 contain programming code or an explanatory text, equations, or figures.
 Data-driven models written in those notebooks rely on a set of programming code
 cells that can take a few milliseconds to a few hours to execute. The execution
 time depends on a variety of factors, including libraries in which a model
 depends on, functions being called, function parameters, algorithmic
 complexity, data set sizes and structures, and, naturally, the computing
 infrastructure responsible for running the code contained in the cells. The
 execution of code and return of results are handled by a back-end component
 called \emph{kernel}, an independent process started by the Jupyter server to
 execute user code.

Users may run these interactive computing notebooks either in their own devices,
such as laptops, desktops, tablets, and smartphones or in a remote computing
infrastructure such as a cloud environment or a supercomputing facility.
Normally, users select a single place to run their notebooks. However, there are
various cases users can benefit from running a notebook in multiple computing
environments: (i) parts of a notebook require processing of data in
a location---e.g. for anonymization or filtering---whereas model training
algorithm requires specialized resources from another location; (ii) data
visualization of intermediate results can happen in a location closer to
the user, while data is processed in a remote location; or
(iii) parts of a notebook may depend on libraries or software stacks that are
ready for consumption in a remote location, which would be more difficult for
a user to setup such complex software environment locally. Those use cases are
triggered by several factors including cell execution time, monetary costs to
access computing resources, privacy issues, data location, data access
restrictions, and specialized hardware and software availability.  All of this
comes with a challenge of deciding which, when, and how cells should be migrated
across computing platforms while keeping a great experience to users working in
their data science notebooks.

This paper addresses the challenge of helping users leverage multiple computing
platforms to run their interactive computing notebooks. Therefore, the
contributions of this paper are: \

\begin{itemize}
 \setlength\itemsep{0.6em}
    
\item Creation of a Jupyter notebook extension to automatically manage the
migration of the \emph{notebook state} of an interactive computing notebook
across computing platforms---requiring no modification to the user source code
($\S$~\ref{sec:jupyterlab});

\item A policy to decide when and how to migrate cells considering the patterns
of user interaction with the notebooks and a set of variables to support such
decision---including cell content, migration costs, and performance
characteristics of the target platforms
($\S$~\ref{sec:context_aware_analyzer}).

\item A solution to reduce the size of a notebook state based on live tracking
of dependencies in programming code cells and abstract syntax trees, decreasing migration time and
improving user experience as a consequence ($\S$~\ref{sec:ast});

\item An extensive set of experiments considering multiple scenarios to
demonstrate the benefits of reducing the notebook states via abstract syntax
trees and the benefits of considering the user context while deciding which
cells should be migrated and when. Experiments are based on
notebooks from Earth science (remote sensing), image recognition, and hand written digit identification ($\S$~\ref{sec:evaluation}).

\end{itemize}

\begin{figure*}[!th]
    \centering
    \includegraphics[width=1.0\linewidth]{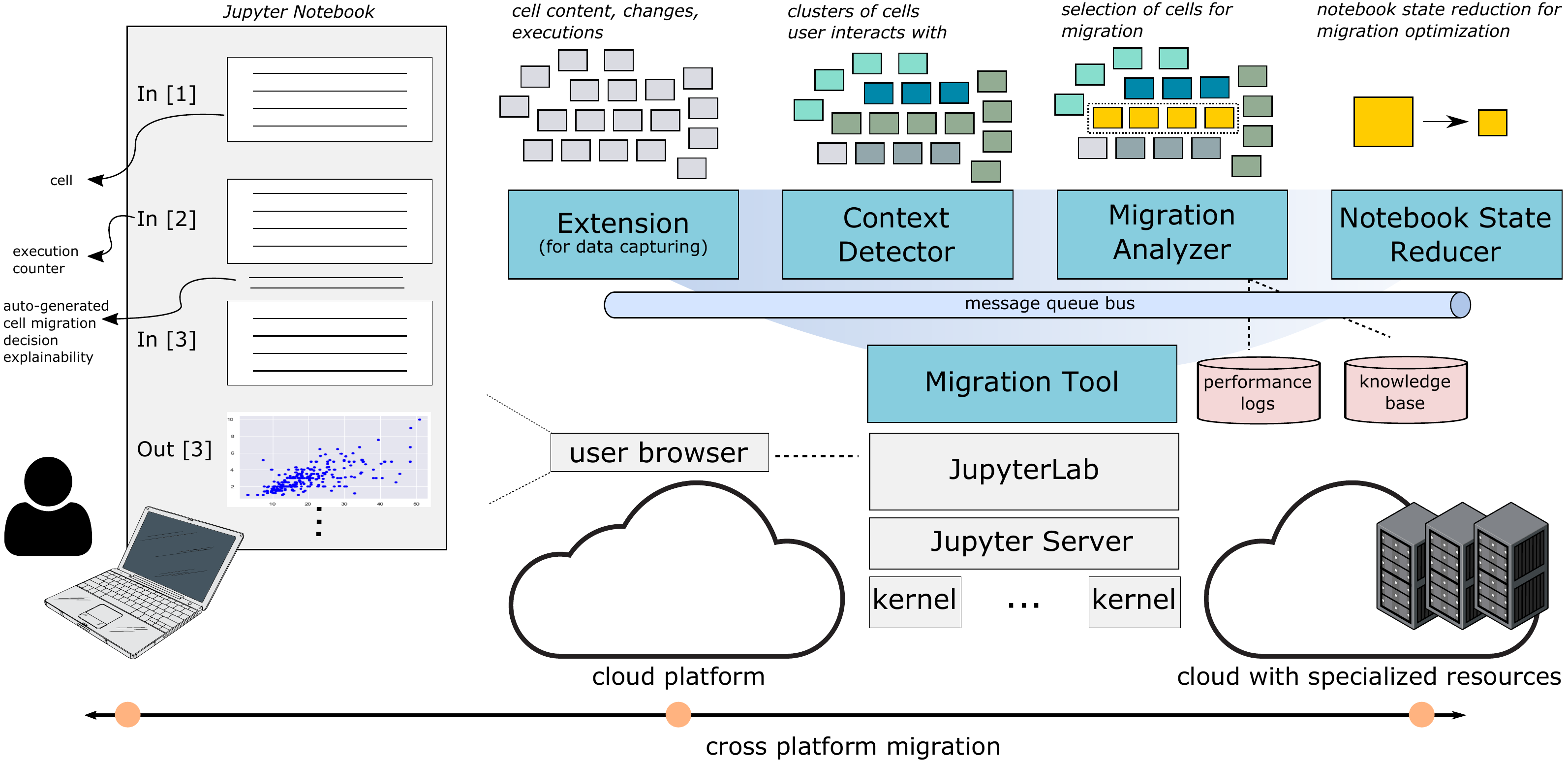}
    \caption{%
        Solution overview, which comprises four major components to enable cell executions across multiple platforms: JupyterLab extension for data capturing while the user is interacting with the notebook, context detector for clustering sequences of cells that a user interacts with, migration analyzer for deciding which cells should be migrated to a remote platform, and notebook state reducer for removing unnecessary data to speed up notebook state migration process. For the migration analyzer, there are two major sources of data this component relies on: one is the performance logs of previous executions of cells in multiple computing environments and a knowledge base to help understand the semantics of cells. Cells are automatically annotated with explainability on cell migration decisions---for instance showing performance gains running in a remote host compared to local environment.
    }
    \label{fig:solutionoverview}
\end{figure*}

\section{Solution}
The proposed solution for users to utilize multiple computing platforms when
using interactive computing notebooks is illustrated in Figure
\ref{fig:solutionoverview} and detailed in the following sections.
\subsection{Extending JupyterLab for Context-aware
Migration}\label{sec:jupyterlab}

For the purposes of this paper, we will focus our discussion on JupyterLab,
a web-based IDE for interactive computing notebooks.  The reader should be
aware that the Jupyter project has different software components, some of them
able to operate on files that follow the Notebook Document Format. An older
project that operates on such files is the Jupyter Notebook application.
Throughout this document, unless explicitly stated, we will be talking about
Jupyter Notebook files when we mention notebooks. When represented in memory,
notebooks are essentially lists of cells, with each cell in the default
implementation being either a code cell (for executable source code), an
attachment cell (for documentation in Markdown), or a raw cell (mostly for
conversion purposes).

JupyterLab also supports a plugin system, enabling users to develop and install
\emph{extensions}. Most of the functionality provided by JupyterLab is
implemented as a set of extensions. We have implemented one such extension as the
core component of our tool, which only operates on code cells and ignores
other cell types. The left part of Figure~\ref{fig:solutionoverview} shows
three code cells with an example of output for the third one.  JupyterLab
extensions can target both the frond-end as well as the back-end of
JupyterLab, with front-end components being written in JavaScript or TypeScript
(which compiles to JavaScript) and executing in the user's browser, while
back-end extensions are written in Python and run in the Jupyter server.
Although the current version of the extension operates in the back-end, it
needs awareness of the user's actions to detect the current context
(\S~\ref{sec:context-detector}) and to decide when to migrate cells
(\S~\ref{sec:context_aware_analyzer}).

For each relevant high-level user action in the front-end, we generate
equivalent telemetry messages that are broadcast to the other components of the
system via the server-side extension, bridging the client side of JupyterLab
to our message-queue (MQ) bus. To reduce the dependencies on the user
side, and to make sure no new networking requirements are added on top of
JupyterLab, our extension creates an authenticated endpoint for telemetry
messages. All well-formed telemetry messages sent to this endpoint are
forwarded to the MQ bus for consumption of other components of the system.

All telemetry messages have a datetime field, which represents the moment at
which the message was created, a reference to a cell id (a UUID uniquely
identifying cells in Jupyter Lab), a reference to the notebook the message is
related to, the list of cell ids currently in the notebook, a UUID identifier
for the current session, the current path of the notebook relative to the
working directory of the Jupyter server, and a message \emph{type}. The various
types we have defined are described in Table~\ref{tab:message-types}.

\begin{table}
  \centering
  \caption{
    Types of telemetry messages used in our extension.
  }\label{tab:message-types}
  \begin{tabular}{lr}
    \toprule
      \textbf{Telemetry message type} & \textbf{Message sending triggers} \\
    \midrule
    \texttt{session-started} & New notebook session is started\\
    \texttt{session-disposed} & Notebook session is closed\\
    \texttt{cell-execution-requested} & User requests execution of a cell\\
    \texttt{cell-execution-started} & The kernel starts to execute a cell \\
    \texttt{cell-execution-completed} & The kernel finishes executing a cell\\
    \texttt{cell-modified} & A cell was modified and lost focus \\
    \bottomrule
  \end{tabular}
\end{table}

The front-end of the extension was implemented by creating listeners
for events that occur in the notebook application. The most important events
are:
\begin{enumerate}
  \item A new user session becoming ready, which tells the
    extension when the user is starting work on a new notebook, allowing
    us to inject code in the user's kernel for transparent migration;
  \item Any changes to kernel state and to session status, so that we know when
    the kernel is working, and whether it was restarted, enabling the detection
    of when code needs to be injected into the user's SessionContext;
  \item Messages exchanged in Jupyter's \IOpub{} channel\footnote{
      The \IOpub{} channel is where the kernel publishes all side effects of
      executing cells, such as standard output/error and debugging
      events.
    }.%
\end{enumerate}

We also exploit the dynamic nature of JavaScript to patch, at runtime,
JupyterLab's function responsible for instructing the kernel to execute code in
Jupyter's \texttt{Shell} channel. This runtime patch is needed because even
though JupyterLab exposes a signal for listening to messages in the
\texttt{Shell} channel, listeners are prevented from modifying
messages. Moreover, they are only notified \emph{after} messages are sent.
Without this patch, we would be unable to modify messages carrying user code,
making the kernel unaware of the various computing environments the user has
access to. With this hook and the set of event listeners mentioned, we are able
to analyze the user's interaction with the system, and to transparently change
user code, without the need for human intervention.

To reduce the amount of modifications needed in each cell, when the extension
detects a new session is started, or when an existing kernel is restarted, it
injects some code (a \emph{preamble}) into the newly-created kernel. The
preamble is responsible for two tasks: (i) initiating connections with
execution engines in the environments the user has access to\footnote{
  Although the user does not have to modify code for execution on remote
  engines, we require the user to set up an IPyParallel cluster, otherwise execution defaults to the machine hosting the Jupyter server.}
and (ii) registering a Jupyter cell \emph{magic} that finds data dependencies
in the code in a cell and, if needed, compresses that data and migrates the
cell. This cell magic is then prepended to every cell by the hook
mentioned in the previous paragraph.

\subsection{Context Detector}\label{sec:context-detector}

This component aims at understanding patterns on how users are interacting with
a notebook. Our current implementation, with pseudo-code shown in
Algorithm~\ref{alg:alg_get_context}, is based on the analysis of the history of
user interactions with a notebook to understand common sequences of cells that
have been modified or executed.

The context detector subscribes to the MQ bus for messages sent by the
extension. It starts by getting all sequences of cells from all user
interactions with the notebook (Line 1). For our purposes, a sequence is a
list of nondecreasing cell ids. For example, $1,2,3,2,3$ contains two
sequences: $1,2,3$ and $2,3$. In short, every time there is a gap between
the next and ongoing cell id order, a new sequence is created.

\begin{algorithm}[t]
  \DontPrintSemicolon
  \begin{small}
  \KwIn{
  
  $cell\_order$: current user cell id order\\
  \algorithmiccomment{order is the position of the cell in the notebook}
  $history\_order$: cell id orders from all past user interactions.
  
  }

  \KwOut{

  $sequence\_stats$: list and scores of cell sequences previously executed containing the current active cell.
  }
  $sequences$ = get\_sequences($history\_order$, $cell\_order$)
  \algorithmiccomment{create new sequence whenever the current one is broken}

 $sequence\_stats$ = createdictionary() \algorithmiccomment{common sequences}\;

$total$ = 0\;
sort\_sequences($sequences$, $bylength=increasing$)\;
 
    \ForEach{$seq \in sequences$}{
    $subtotal$ = 1\;
        \ForEach{$seq\_other \in sequences$}{

                        \If {$seq \neq seq\_other$}{
                            
                        \algorithmiccomment{duplicate sequences are removed with subtotal also incremented}\;
                        \If {$seq subset \subset seq\_other$}{
                            $subtotal$  += 1\;
                            
        }
        }

    }
    $sequence\_stats[seq]$ = $subtotal$\;
    $total$ += $subtotal$\;
    }
  
    \ForEach{$key,score \in sequence\_stats$}{
        $sequence\_stats[key]$ = $score$/$total$ * 100 \;
    }
   
   return $sequence\_stats$\;
  \end{small}
 \caption{Get context from notebook interactions.}
 \label{alg:alg_get_context}
\end{algorithm}

\begin{figure}
    \centering
    \includegraphics[width=0.8\linewidth]{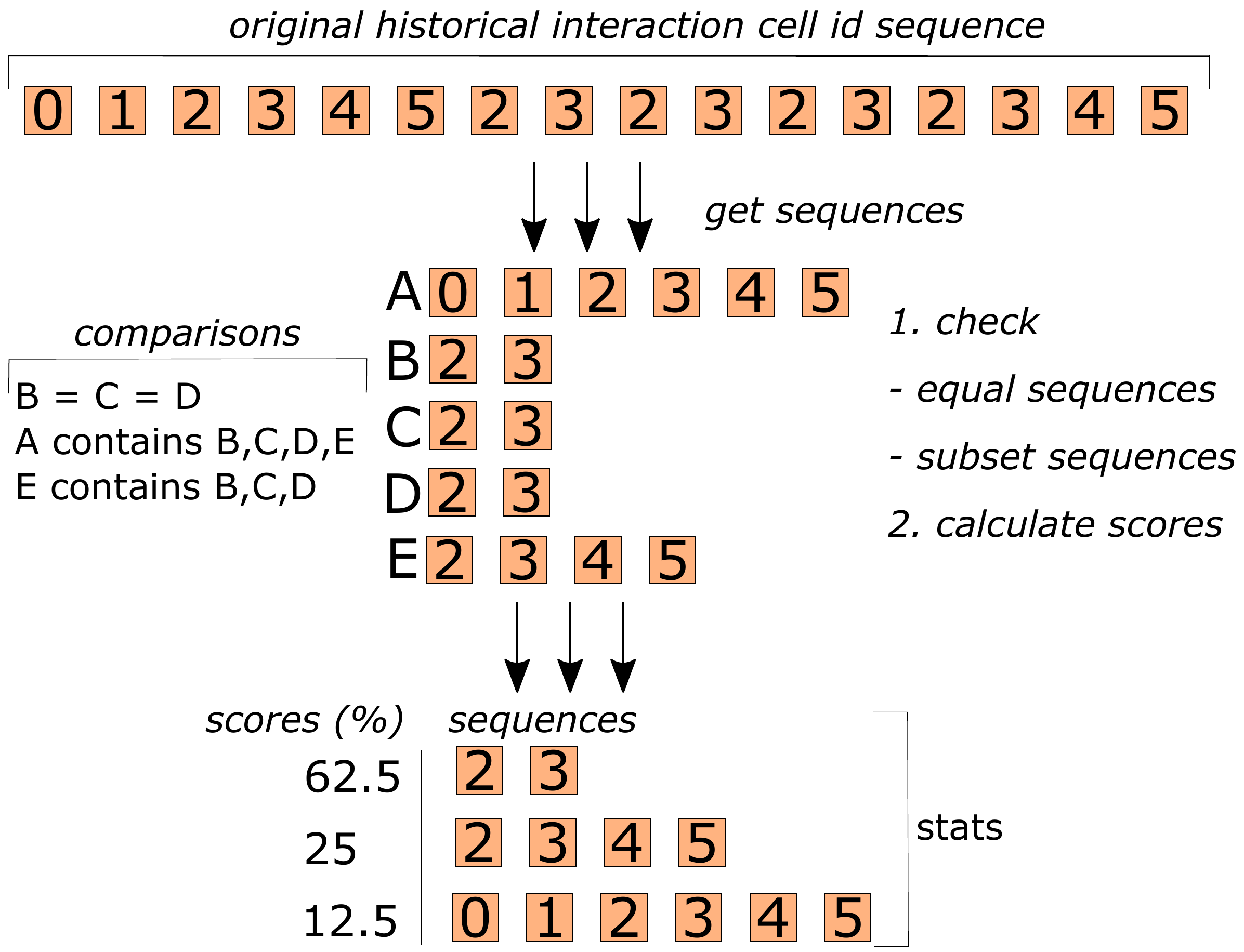}
    \caption{Example of a sequence and the contextual statistical data
        generated by the context detector.
    }
    \label{fig:example_context_sequence}
\end{figure}

With all sequences identified, the algorithm scores them by verifying
which are duplicate and which are subsequences. The score is a percentage
that measures how frequently a sequence occurs in the interaction
history. These statistics are stored in a dictionary (Line 2), filled in
increasing order of number of elements (Line 4), with duplicates removed, but
counted accordingly (Lines 7--11, $subtotal$ variable). Lines 14--15 normalize
the computed values before returning the final result.
Figure~\ref{fig:example_context_sequence} illustrates major steps of the
algorithm using an example of sequence history.

\subsection{Context-aware Migration Analyzer}
\label{sec:context_aware_analyzer}

The migration analyzer decides if a cell/block of cells should migrate to a
remote infrastructure using knowledge- and performance-aware policies. The first
makes decisions based on previous knowledge about the cells' contents, whereas
the latter relies on previous cell executions. This component uses a background
process for listening to the MQ bus and a Knowledge Base (KB) to support
knowledge-aware migrations. It also uses the context detector
($\S$~\ref{sec:context-detector}) to detect a group of cells to migrate in
blocks. 

\medskip
\noindent \textbf{Knowledge-aware policy}. This policy relies on a KB that
contains cell parameters and estimates which combinations of values make the
computational costs worth a cell migration to a remote infrastructure. Examples
of such parameters (also known as hyperparameters in Data Science) are
\emph{number of epochs, batch size, train/test split size}. For the initial
system state, the estimate values can be hand-crafted by a data science expert,
but as the user writes and executes cells in the Notebook, these estimates are
dynamically updated.  This component contains a ``Notebook to Knowledge Base'' service to retrieve
cell parameters and values by parsing the cell's content using
ASTs. This process generates a data structure following PROV-ML~\cite{souza2020workflow},
an efficient provenance-based Machine Learning and Data Science ontology that extends W3C
PROV-O~\cite{provw3c}, sends the structure back to the decider, and stores the
structure in the KB for provenance purposes. Suppose, for instance, that the
estimated $epochs$ value $e$, for which a local training should be migrated, is
known (is stored in the KB). If the user attempts to train a model with
a number of epochs greater than $e$, the decision is to migrate the cell based
on this criterion.

\medskip
\noindent  \textbf{Performance-aware policy.} This policy uses estimates of
cell execution time in the local infrastructure, the migration time, and the
speedup of running in a remote infrastructure. If the predicted execution time for the
current cell on the remote infrastructure is less than the local execution time
(plus the migration times), the current cell should be migrated. A cell can be
migrated using single-cell or block-cell approaches.

The \emph{single-cell migration approach} is designed for migrating single
cells without considering their execution behavior. In this case,
computationally intensive single cells are executed remotely, triggering the
exchange of notebook state from local to remote and back, when the cell
execution is over.

\begin{figure}
    \centering
    \includegraphics[width=.7\linewidth]{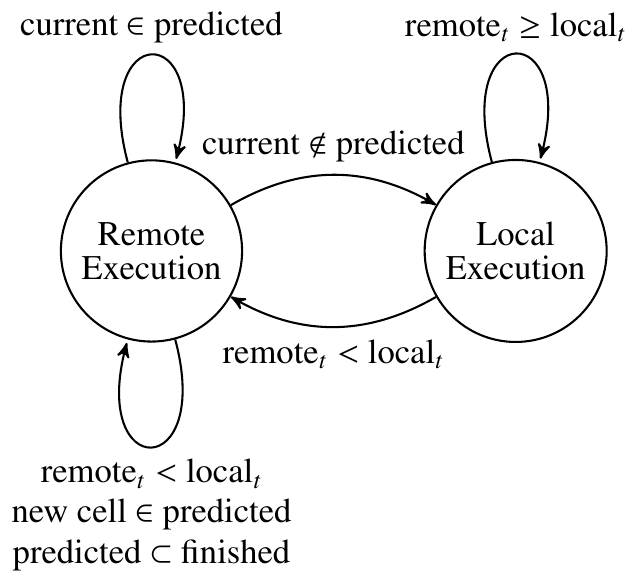}
    \caption{%
        Block migration strategy. The subscript $t$ indicates the time it takes
        to execute a cell in a given system. local$_t$ indicates the run time
        in the local system, and remote$_t$ indicates the run time in the
        remote system.
    }\label{fig:block-cell-migration}
\end{figure}

The \emph{block-cell migration approach} leverages user notebook interaction
patterns to make intelligent cell migration decisions. Unlike the previous
migration strategy, in which each remotely executed cell costs two data
migrations between local and remote hosts, this approach makes migrations only
when necessary according to the users' utilization behavior.
Figure~\ref{fig:block-cell-migration} details the block-cell migration approach.

Groups of cells are migrated to remote hosts whenever the context detector
predicts groups of cells are about to be executed. Therefore, once the
cell-related data migrates to the remote infrastructure it is only switched
back to local execution in two situations: (i) when the user executes all the
cells in the predicted cell group or (ii) when user runs a cell not featured in
the predicted sequence.

\medskip
\noindent  \textbf{Dynamic migration parameters update}. The parameter value
estimates used by the knowledge-aware policy are continuously updated in the
background during the notebook utilization.
Algorithm~\ref{alg:alg_parameters_update} presents the way they are updated.
For any cell event triggered (Line 3), the algorithm iterates over the notebook cells
(Line 4) and gets the structured knowledge from the
cell content (Line 5), verifying whether a cell in the notebook contains
a parameter the KB knows how to handle, likely triggering a migration (Lines
6--7).

Using our model training example, we define as \emph{cell of interest} the
one that fits the model passing hyperparameters to the \emph{fit} method, as
it is known beforehand that this method often requires heavy computation. If a
cell of interest is identified, the algorithm calls the migration manager, which
implements the \emph{build\_or\_update\_dataset()} method that builds the
training dataset used by this algorithm. In this method's initial state, the
migration manager proactively executes the cells in background locally with the
parameters of interest (i.e., the ones that likely trigger a migration) using
values that make the cell execute fast.

In the \emph{model.fit(epochs=10)} cell in our example, the migration manager obtains
 summarized statistics by executing that cell a few times with a varying number
 of small epochs (e.g., 1, 5) in both environments. After both local and
 remote jobs finish, the execution times for parameters values that do not
 require a long time to execute are recorded into a dataset.

In this first implementation, we define a maximum time for executing both jobs
(transparently executed in parallel). As this happens in the
background, users only notice this process if they try to execute cells of
interest before these jobs finish.

After the initial state, this method updates the dataset if the users change the
notebook in a way that impacts the migration time or if they execute a cell of
interest. With this dataset, we train two ML models (one for the local times and
another one for the remote times) to predict the behavior of the execution times
for parameter values that demand long waiting times.

Finally, the estimated value for a given parameter, notebook, and environment
is given by the intersection of the predictions of the local and remote
models (Line 12), with the KB being updated for usage by the knowledge-aware
policy (Line 13).

\begin{algorithm}
  \DontPrintSemicolon
  \begin{small}
  \KwIn{
  
  $MQ$: The events queue with messages from the Extension.
  
  $KB$: The Knowledge Base.
  
  }

  $manager$ = get\_migration\_manager()  \;

  $known\_parameters$ = $KB$.get\_known\_parameters()

  \While{$msg = MQ.listen()$}{
  
        \ForEach{$cell \in msg.cells$}{
        
            $cell\_kwg$ = notebook\_to\_kb(cell)
            
            \ForEach{$p \in known\_parameters$}{
        
                \If {$p \in cell\_kwg$}{
                        $dataset$ = $manager$.build\_\\or\_update\_dataset($cell$, $cell\_kwg$, $p$)\;
                        
                        \If {$dataset.was\_modified()$}{
                            $m\_local, m\_remote$ = train($dataset$)\;
                            
                            $opt\_val$ = intersection($m\_local$, $m\_remote$) \;
                            
                            $KB$.update($p$, $opt\_val$)
                        }
                     
                }
            
            }

        }
   }
   
  \end{small}
 \caption{Dynamic migration parameters update.}
 \label{alg:alg_parameters_update}
\end{algorithm} 

\subsection{Notebook State Reducer}\label{sec:ast}

Traditional software, passed as a single source code unit to a compiler,
benefits from certain optimizations that interactive computing notebooks do
not.  For instance, compilers can detect when variables with global scope are
no longer needed, producing bytecode that garbage collects them at an
appropriate time. In an interactive computing environment, where the user
constantly modifies the contents of cells, possibly going back to cells
previously executed to re-run them one more time, such optimizations do not
apply---it is not possible to tell in advance which variables and functions
will be accessed by a cell in a future execution.

Because memory resources may not be automatically freed as a developer of
traditional software would expect, the memory footprint of the runtime
environment can grow very large. In a scenario where the notebook state is
migrated to execute on a different computing platform, this is undesirable, as
serializing objects will take more time to complete, causing data to take
longer to transfer over the network.  This problem is exacerbated if, at some
point, the notebook state needs to be transferred back from the remote host to
the original one.

The aforementioned issues have led us to design a method to identify and
eliminate unneeded objects from the notebook state as it is serialized. Our
method relies on the analysis of the source code embedded in the cell that
has been marked for remote execution by the Migration Analyzer component. 

We begin by processing the source code of a cell with an abstract syntax tree
(AST): a tree of objects that includes expressions, literals, statements,
control flow elements, and variable references. Data loads, designated by
a \emph{Load} node connected to a name, are the primary source we use to
identify the dependencies of a cell. When such a node is found, we search
the global namespace for the associated name to identify if it relates to
a variable or a function and mark that object as ``needed'': the execution
of that cell depends on it.

When a variable object is marked as such, the method recursively inspects
the definition of that variable to build a data dependency graph,
allowing, e.g., arrays holding references to other objects to have all their
dependencies properly captured. When the object refers to a function, the method
registers that function as ``needed'' and proceeds to a recursive inspection to
identify other functions and global variables that have to be incorporated into
the final dependencies list. The method also scans loaded modules for
dependencies.

Once there are no more objects to be processed, all objects linked to the
notebook state that do not belong to the dependencies list are temporarily
detached from that state. This allows the serialization code to operate on
a reduced set of data that represents the exact requirements that have to be
transferred to the remote host so that the cell can execute to completion.
The objects are attached back once the serialization process completes.
In the event of a serialization failure, the cell executes locally.

An advantage of doing dependency analysis at run time (in comparison to
static analysis) is that no time is wasted processing code branches that
the current execution did not take. Also, it captures dynamic processes
influenced by user input, such as the inclusion or removal of objects from
arrays.

We also employ the following steps to reduce the size of the serialized
notebook state even further:

\begin{itemize}
\item First migration from local to the remote host: the
    notebook state as described above is captured and transferred;
\item Migration from remote back to the local host: only
    newly created objects or objects that have been modified (i.e., whose
    hashes have changed after the cell has executed) are serialized
    and transferred to the target. Unhasheable objects are always
    migrated -- at the expense of potentially transferring more
    data over the wire than needed;
\item Subsequent migrations from local to the remote host:
    only serialize the differences as described in the previous item.
\end{itemize}

It is also possible to apply compression to the serialized data to decrease
network transfer times (at the cost of compression and decompression times,
which may vary according to the algorithm of choice and its implementation).
A discussion of compression algorithms is beyond the scope of this paper,
although we do provide some numbers in Section~\ref{sec:evaluation}.
\section{Evaluation}\label{sec:evaluation}

This section presents two sets of experiments to investigate efficiency of the
proposed migration tool. In the first set, we show results related to the
notebook state migration process for a given set of notebook cells. We
investigate the performance of migration in terms of memory consumption and
migration time. In the second set of experiments, we present results on the
interactive notebook execution using different strategies for cell migration
ranging from local to remote execution.

\subsection{Notebook State Reducer for Migration}

For this experiment we use a Jupyter notebook that
processes a collection of satellite scenes from the Spacenet7 challenge
~\cite{spacenet7}. That dataset contains RGB mosaics of 60 regions in the
world, each of which holding a time-series comprising 24 images collected
over a time span of two years. The spatial regions cover areas of 4km x 4km
(represented by grids of 1024x1024x3 pixels) that have seen active urban
development over that time period.
The notebook loads all images from 30 regions of that dataset (i.e., 720
images). The data processing pipeline normalizes and computes the histogram
of each file. The Wasserstein distance of adjacent histograms is calculated
next. Scenes whose resulting distance are below a given threshold are
considered too similar for further processing. In our experiment a total of
93 distinct images are above that selected threshold.

The images from the final set are then handled by a Sobel filter 
that enhances their contours---which reveal both urban structures such
as roads, neighborhoods, and natural structures like water bodies.
Last, each filtered image is clustered into 4 bins using K-Means and converted
into a vectors (shapefile) format. This is the most computing intensive cell
and the actual one chosen by the Migration Analyzer for executing in a more powerful remote resource.
The size of the notebook state captured by our engine is shown in Table~\ref{tab:squeezesize}. Four different configurations are used. In the first,
the full notebook state is captured (i.e., without pruning unneeded objects
first). A variation that uses ZLib compression is also included for comparison.
The remaining configurations are the reduced notebook state we implemented
and its ZLib-compressed variation. Our technique reduces the original notebook
state by 55$\times$ with compression and 8$\times$ without.

\begin{table}
  \centering
  \caption{Notebook state sizes}
  \label{tab:squeezesize}
  \begin{tabular}{llr}
    \toprule
    \textbf{Direction} & \textbf{Capture method} & \textbf{State size} \\
    \midrule
    Local host to remote & Full state &                17,468 MB \\
    Local host to remote & Full state, compressed &     2,655 MB \\
    Local host to remote & Reduced state &              2,231 MB \\
    Local host to remote & Reduced state, compressed &    320 MB \\
    Remote to local host & Full state &                21,932 MB \\
    Remote to local host & Full state, compressed &     4,307 MB \\
    Remote to local host & Reduced state &              4,463 MB \\
    Remote to local host & Reduced state, compressed &  1,652 MB \\
    \bottomrule
  \end{tabular}
\end{table}

Once computation on the remote host is complete and execution of the next cell
is triggered, the Migration Analyzer decides to run it on the local host. Our
cell reducer algorithm identifies new and changed objects in the remote
notebook state and only transfers those back. As shown in
Table~\ref{tab:squeezesize}, our approach reduces the data size by 
13$\times$ (compressed) and 5$\times$ (uncompressed) when compared to the
full notebook state.

\subsection{Context-aware Migration}

\noindent \textbf{Experiment setup.} We assessed four migration policies to
verify the solution feasibility and speedups: (i) \textbf{Local execution:}
baseline without migrations at any time, in which all cells are executed
locally; (ii) \textbf{Single-cell migration:} designed for migrations of single
cells without considering the cell's execution behavior; (iii)
\textbf{Block-cell migration:} leverages understanding of context, i.e. common
executed cells by the user to make intelligent cell migration decisions; (iv)
\textbf{Remote execution:} considers the situation where all cells execute in
a remote host. In this case, we take full advantage of the remote
infrastructure, however this can come with higher execution costs. For
demonstration purposes, in this set of experiments, we forced a fixed migration
time and remote speedups. We assess single-cell and block-cell migrations using both
\textbf{performance-} and \textbf{knowledge-aware} policy variations.

\begin{figure}
  \centering
  \includegraphics[width=\linewidth]{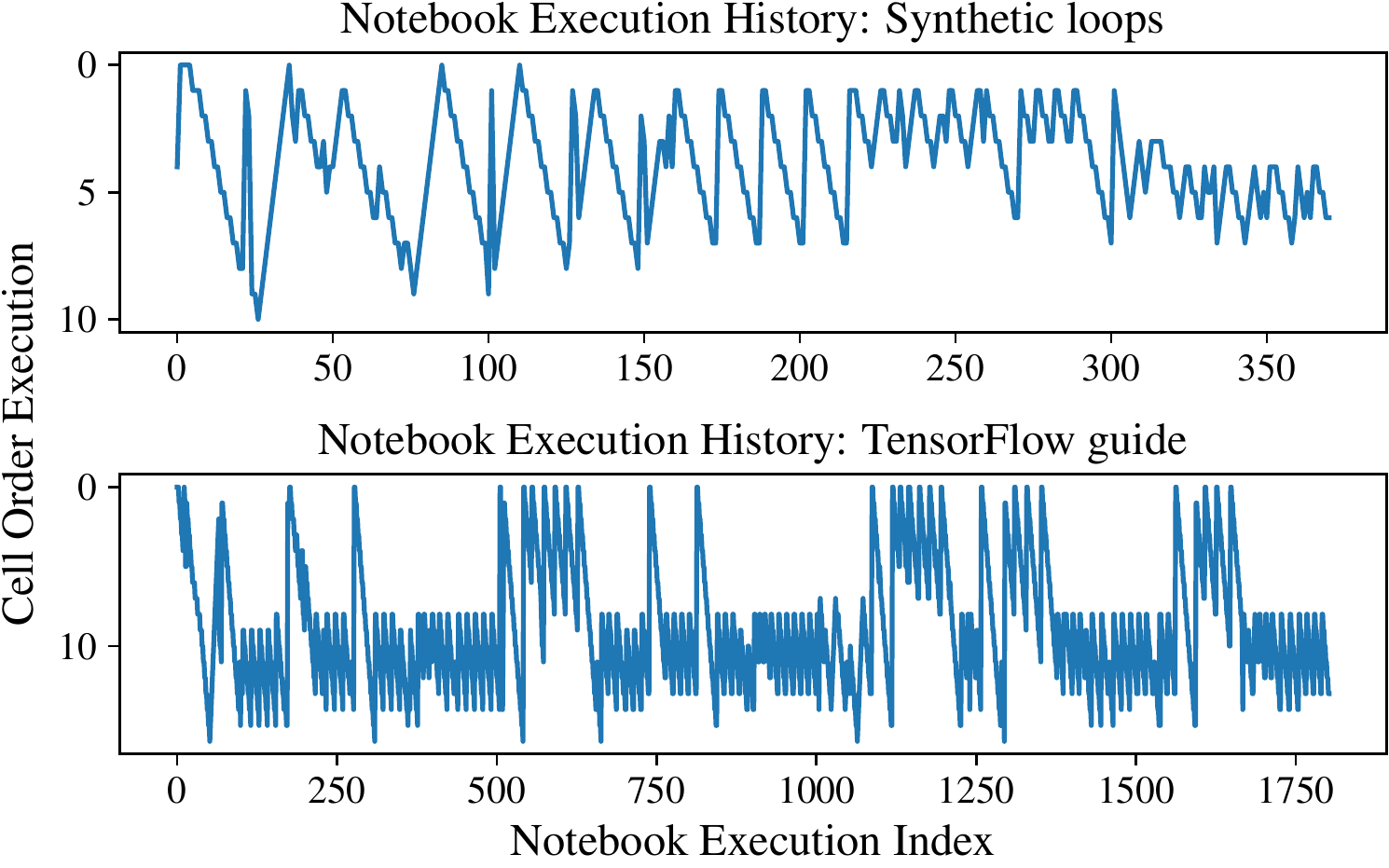}
  \caption{%
    User interactions for the two proposed notebooks in this study: synthetic
    loops and adapted TensorFlow guide. The Notebook Execution Index shows
    the current user step in the notebook utilization. The Notebook Cell
    Order Execution presents the order of the current executed cell.
  }\label{fig:execution_flow}
\end{figure}

For the performance-aware migration policy, we assess two Jupyter notebooks
to evaluate how user interaction context and migration policies impact user
development time. The first notebook is composed of a set of synthetic loops
to emulate heavy computation cells. The second notebook is an adaptation
of the TensorFlow beginner's tutorial and compares different
hyperparameters settings for MNIST \cite{deng2012mnist} dataset
classification. We used the notebooks
throughout the proposed system and depict their cell execution flows in
Figure~\ref{fig:execution_flow}. The graphs show the Notebook execution flow
for different cell orders. In both cases, the reader can see cycles in the
execution flow. For instance, in the synthetic loops notebook the user
executed cells 1--7 several times from notebook Execution Indexes 160--230.
This cell execution behavior is captured by the context analyzer to
improve the cell migration process. For the two notebooks and their
related execution records (Figure~\ref{fig:execution_flow}), we evaluated
four migration policies varying the migration time and remote infrastructure
speedups. 

\medskip
\noindent \textbf{Hardware and software setup}: the local computing
environment comprises a machine with 6 physical CPU cores with 2-way simultaneous
multithreading (12 logical cores), 16 GB RAM DDR4, 512 GB SSD\@.
For the remote environment, we use IBM Research's Cognitive Computing 
Cluster using a computing node with 56 CPU cores, 1.5 TB RAM, a GPFS with up to
296 TB, and an NVIDIA K80 GPU with 4992 CUDA cores and 24GB of GDDR5 vRAM.
As for software, the microservices and libraries are developed in Python, 
the message queue broker is Redis 6, and the NotebookToKB service uses ProvLake
\cite{souza_efficient_2019} as its provenance system. The latter stores the
provenance data as a knowledge graph in the KB that uses the IBM Hyperknowledge
technology \cite{moreno2017hyperknowledge}. The DL training notebook uses
Tensorflow 1.15 and Keras 2.3.1.

\subsection{Result Analysis}

In this section, we assess the impact of the two policies implemented in
the Context-aware Migration Analyzer ($\S$~\ref{sec:context_aware_analyzer}),
i.e., single-cell and block cell migrations, and their variations, i.e. the
performance- and knowledge-aware policy variations. All compared with full
remote and full local execution.

\begin{figure}
  \centering
  \includegraphics[width=\linewidth]{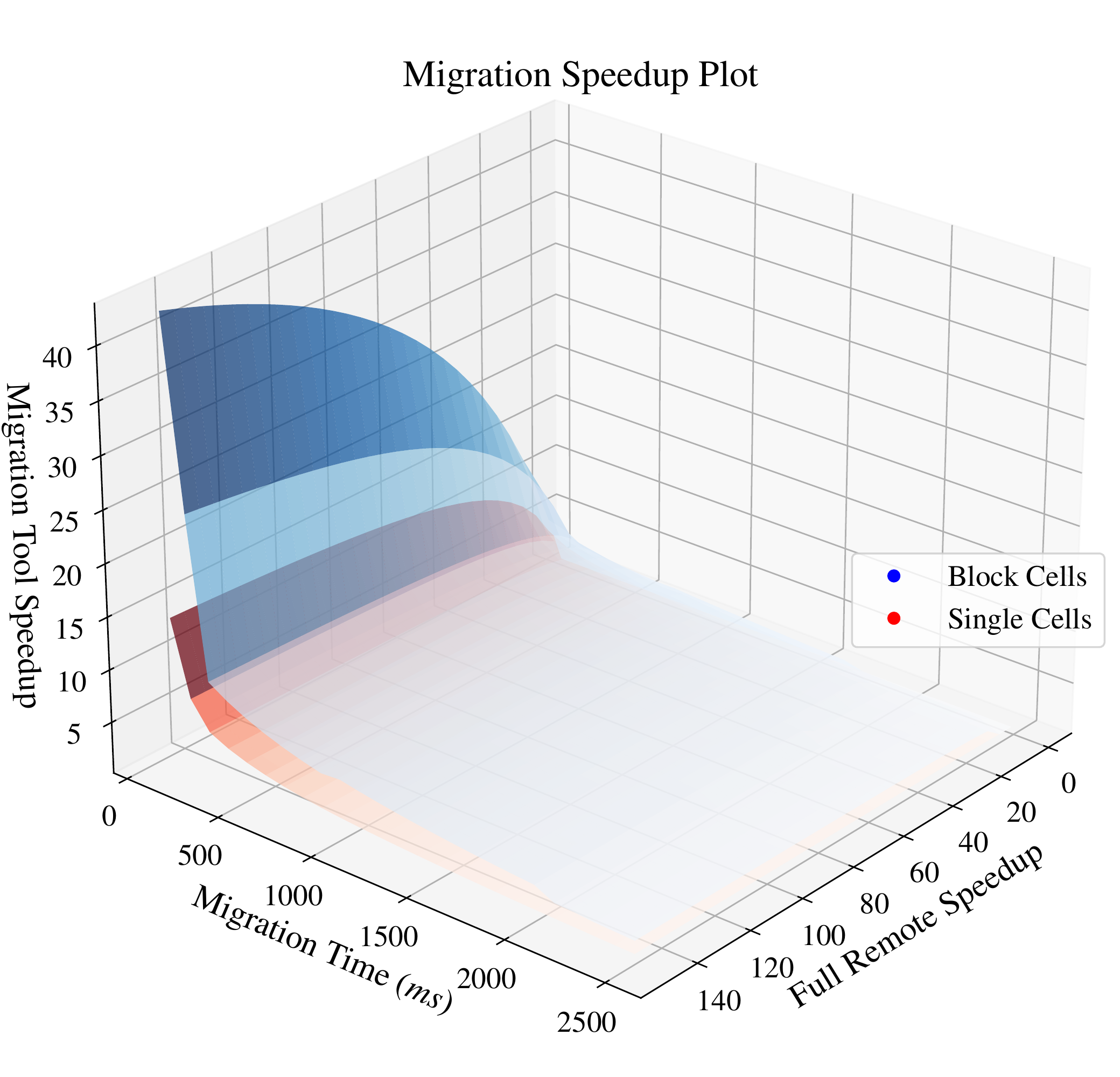}
  \caption{%
      Speedup plot for synthetic loops notebook. This graph presents the
      speedup for block-cell and single-cell migration approaches
      considering different values of migration time and remote
      infrastructure speedups. Readers can see the block-cell migration
      policy outperforms the for all cases and the most significant
      migration speedups happen with low migration times and high values of
      remote speedups.
      }\label{fig:3dplot_1}
\end{figure}

\begin{figure}
  \centering
  \includegraphics[width=\linewidth]{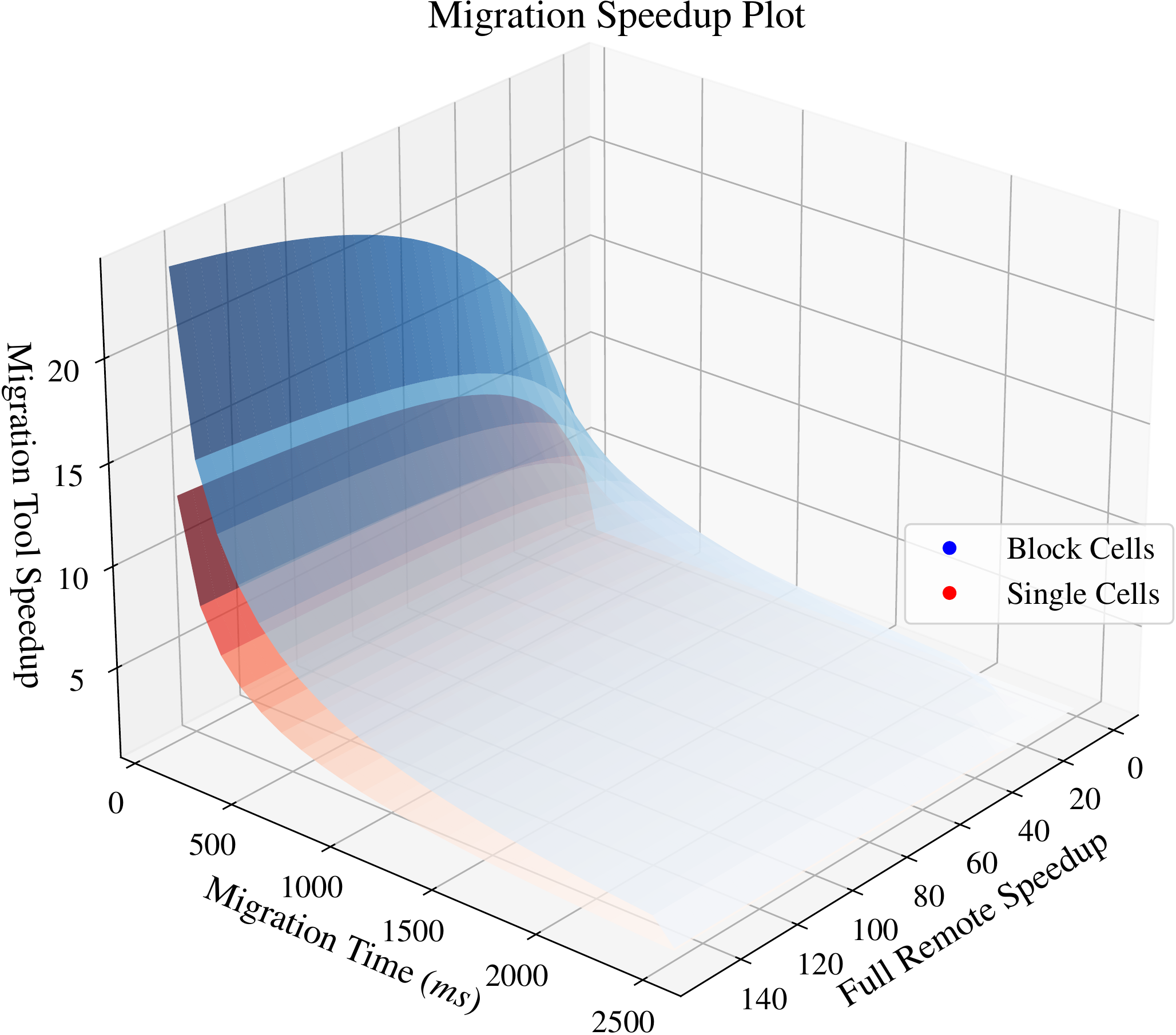}
  \caption{%
      Speedup plot for TensorFlow guide notebook. This graph is similar to
      the synthetic loops notebook with the difference in terms of speedup of
      block-cell migrations. In this case, there are more cell migrations as
      the user creates shorter blocks of executed cells (Figure~\ref{fig:execution_flow}), then, the speedup is less
      significant.
      }\label{fig:3dplot_2}
\end{figure}

\medskip \noindent \textbf{Performance-aware policy.} Figures \ref{fig:3dplot_1}
and \ref{fig:3dplot_2} show the performance migration speedups for block-cells
and single-cell migration strategies using different cell migration times and
full remote speedups. We assumed fixed local and remote infrastructures. The
proposed solution can migrate workloads for different platforms, and the impact
of migration times and remote speedups would vary according to hardware
specifications.  For both graphs, we observe the block-cell
migration strategy outperforms the single-cell based for all combinations of
full remote speedups and migration times. This increased speedup can be
explained as the block-cell migration strategy reduces the number of cell
migrations by leveraging the user cell utilization context. The two plots show
similar shapes both for block-cell and single-cell migration approaches. The
maximum speedup value happens with the minimum migration time and highest full
remote speedup. The migration speedups decrease as long as the full remote speed
decreases and/or the migration time increases.

\begin{figure}
  \centering
  \includegraphics[width=\linewidth]{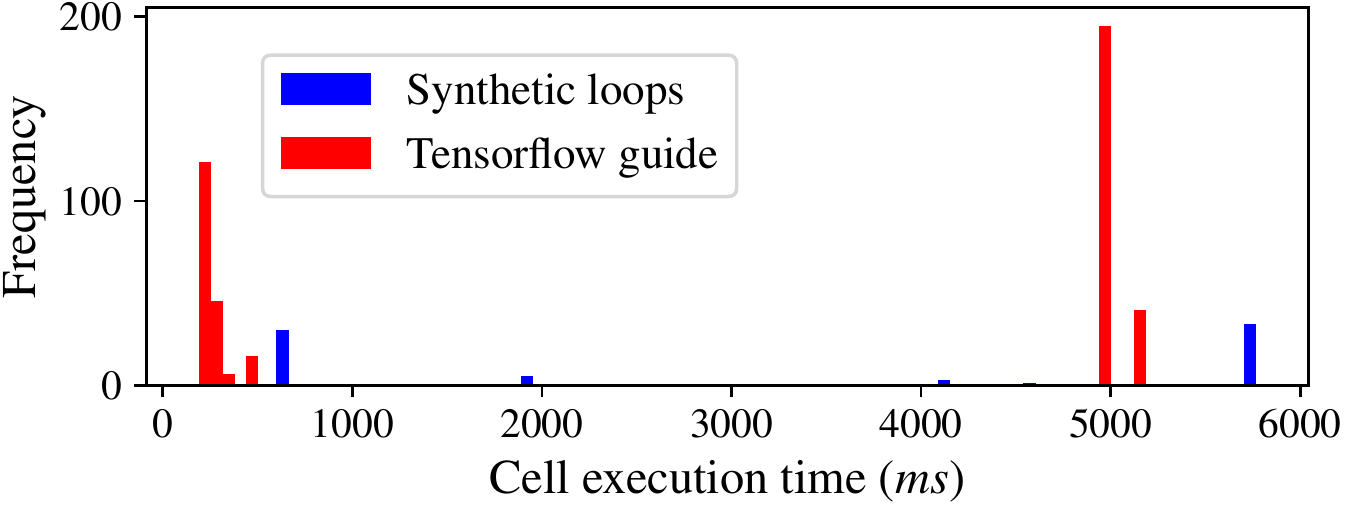}
  \caption{%
      Cell execution times frequency graph. It shows the execution count
      and the relative frequency for executed cells in Synthetic loops and
      TensorFlow guide notebooks.
      }\label{fig:execution_times}
\end{figure}

Single-cell migrations present similar speedups for low migration times in both
plots (see the dark red areas in Figures~\ref{fig:3dplot_1} and
\ref{fig:3dplot_2}), however, the block-cell migration speedups are higher in
the synthetic loops notebook (dark blue regions). There are two explanations for
this: (i) the synthetic loops notebook presents bigger execution cycles when
compared to the adapted Tensorflow guide notebook (Figure
\ref{fig:execution_flow}); this user cell execution behavior forces fewer
migrations as the proposed solution executes larger blocks of cells, and (ii) the
frequency of less computationally intensive cells is higher in the adapted
Tensorflow guide notebook (Figure \ref{fig:execution_times}), making
migrations to the local infrastructure more frequent. 

\begin{figure}
    \centering
    \includegraphics[width=\linewidth]{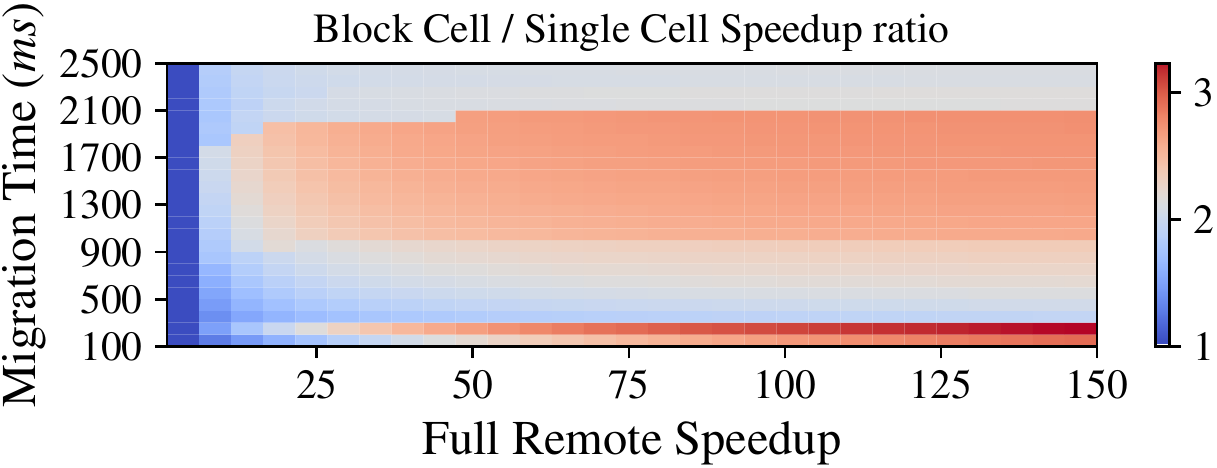}
    \caption{%
      Block-cell/Single-cell speedup ratio for different values of migration
      times and remote infrastructure speedups related to synthetic loops
      notebook.
        }\label{fig:speedup1}
  \end{figure}

\begin{figure}
    \centering
    \includegraphics[width=\linewidth]{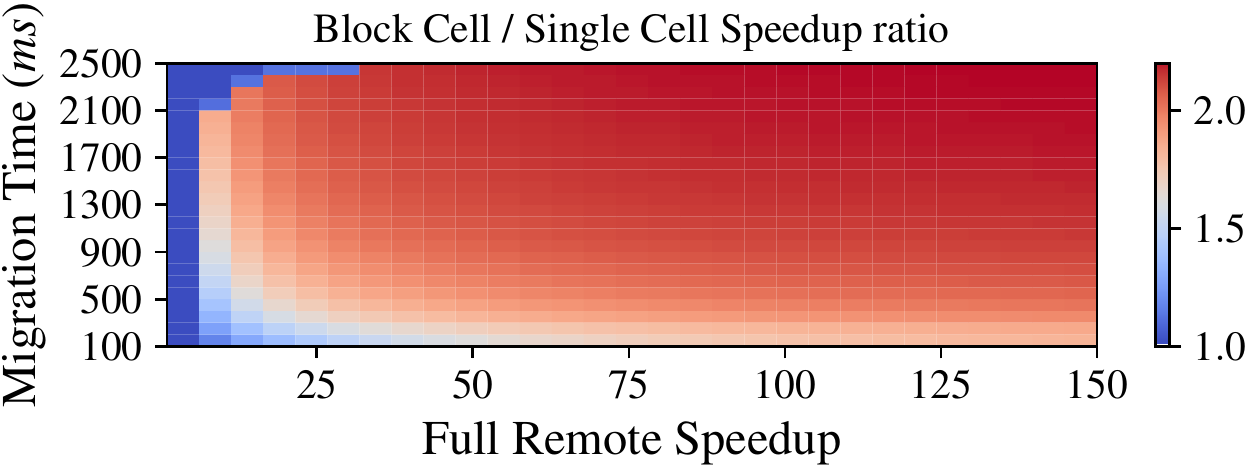}
    \caption{%
      Block-cell/Single-cell speedup ratio for different values of migration
      times and remote infrastructure speedups related to TensorFlow guide notebook.
        }\label{fig:speedup2}
    \end{figure}

\begin{figure}
    \centering
    \includegraphics[width=\linewidth]{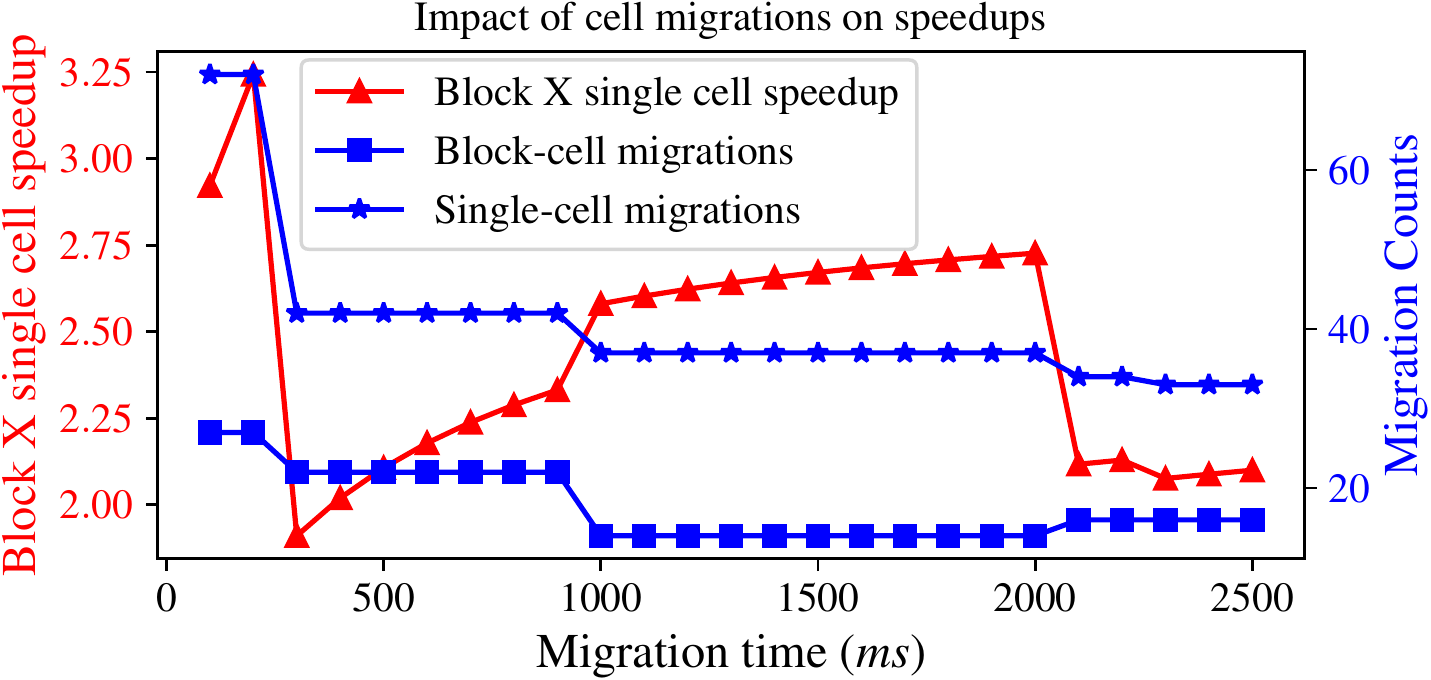}
    \caption{%
    Impact of migration counts on block-cell/single-cell speedup ratios. We
    fixed the remote speed up to 150 times. Then, when the migration counts stay
    the same, the ratio keeps increasing with the rise of migration time.
    However, when the ratio changes, the relationship of Block X Single-cell
    speedups also changes.
    }\label{fig:explanation}
  \end{figure}

  Figures~\ref{fig:speedup1} and \ref{fig:speedup2} show the ratio between the
  speedups using block-cell and single-cell migration approaches for the two
  presented notebooks. For both figures, the speedup ratio is close to one when
  the full remote speedup is small and rises as the speedup increases. In other
  words, the block-cell migration approach takes more benefit from full remote
  speedups when compared to the single-cell migration approach. When we consider
  migration time increasing, the previous conclusion is not the same for both
  notebooks. For the adapted TensorFlow guide notebook, scenarios where the
  migration times are higher, the block-cell migration approach performs better
  when compared to the single-cell method. It happens as the block-cell
  migration approach performs fewer migrations than the single-cell migration
  method. For the synthetic loops notebook, migration time increases do not reflect
  directly into rises in the block-cell/single-cell speedups ratio as we would
  first infer. To explain this phenomenon, Figure~\ref{fig:explanation}, which presents a slice of Figure \ref{fig:speedup1} with full remote speedup equals 150, shows the impact of migration
  counts for block-cell and single-cell migration policies on its speedups
  ratio. One can see when the proportion of block-cell and single-cell
  migrations remain constant (e.g., migration time between 1000 and 2000
  milliseconds) the speedup ratio keeps increasing with the migration time
  increase, just like Figure \ref{fig:speedup2}. However, for this notebook,
  the cell execution times are more scattered when compared to the adapted
  TensorFlow guide notebook, which contains two main groups of execution times
  (Figure \ref{fig:execution_times}). These different groups of cell execution
  times change the number of migrations for both migration policies, altering
  the effect of migration time increasing on the speedup ratio (e.g., see
  migration time change from 900 to 1000ms).
  According to the results, we can see the following: (i) knowing the user cell
  execution flow can improve the migration policies and maintain a good
  compromise between execution costs and computational power during the
  interactive notebook construction, and (ii) the block-cell migration approach
  performs better than the single-cell migration policy as it leverages the user
  execution flow to improve the data exchange process.

\begin{figure}
    \centering
    \includegraphics[width=1.0\linewidth]{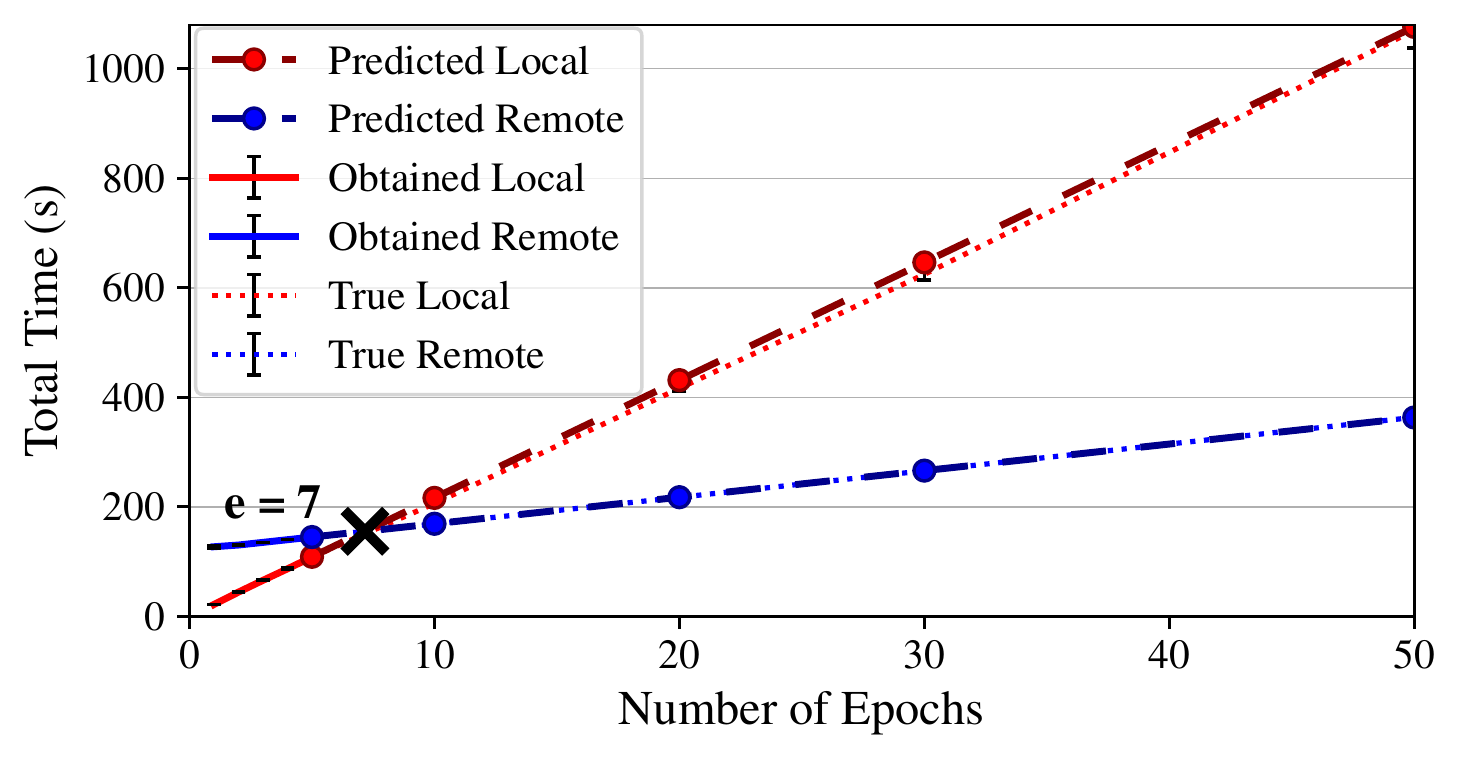}
    \caption{%
        The algorithm automatically finds that for epochs $e>7$, 
        the migration time pays off to migrate to an environment that runs 4.43x faster.}
    \label{fig:content_aware_epochs}
  \end{figure}

\medskip \noindent \textbf{Knowledge-aware policy.} To evaluate this policy, we
use a Deep Learning (DL) training Notebook as a motivating use case. In this
example, the data scientist trains an image classifier DL model using the
Cifar100
dataset\footnote{https://www.machinecurve.com/index.php/2020/02/09/how-to-build-a-convnet-for-cifar-10-and-cifar-100-classification-with-keras/}.
In this evaluation, we assess the hyperparameter number of epochs for a model
training using the TensorFlow Keras library. In the initial state, the KB
contains the information that, for the Cifar100 dataset, an estimate value for
the number of epochs is $e=50$, which is manually informed by an expert
(including the valid ranges).  Nevertheless, $e=50$ may be sub or
superestimating the optimal epoch value to migrate. For this reason, Algorithm
\ref{alg:alg_parameters_update} runs in background to build the training
datasets with small candidate values for $e$. In this test, we set the migration
time to 2 minutes and the maximum waiting time to 5 minutes, meaning that this
is the time limit to build the initial dataset after the Migration Analyzer
service identifies that a model is going to be fitted in the user's Notebook.
With the hardware in use in the experiments, the small number of epochs that can
execute both locally and remotely while satisfying these constraints (and with a
minimum amount of repetitions to analyze the standard deviation) is $\{1, 2,
3\}$. The times obtained for the local and remote environments for these
numbers of epochs are used by the Migration Analyzer to train the local and
remote ML models, using linear regression (see
Fig.~\ref{fig:content_aware_epochs}).

The red line represents the local times, whereas the blue line represents the
remote times. The solid lines represent the obtained times for the small numbers
of epochs $\{1, 2, 3\}$ executed and that were used as the training datasets for
the regressors. To assess the accuracy of the predictions for greater numbers of
epochs, we also execute these greater values to get their true total times
(dotted lines in the plot). We expected that the predictions made by the
regressors would have a good fit with the true lines, since the number of epochs
is directly proportional to the training time and we are leaving all other
hyperparameters constants. For this reason, the results confirm that the linear
regressors are a simple and unexpensive solution to predict the total times for greater
numbers of epochs and we also find that this holds even for a very small
training dataset (3 obtained samples). The obtention of the total times repeated
while the standard deviation of a minimum of two measurements was below 10\% of
the median. The obtained and true lines represent the median of these
repetitions. We cannot repeat each measurement many times because each
repetition counts for the user waiting time. Still, we see the prediction lines
close to the true lines and the error bars, which represent the standard
deviations of the repetitions, are very small. Furthermore, the lines for the
remote times begin in a higher total time due to the migration time from the
local to the remote host. However, since the remote host has a much more
powerful hardware, particularly a GPU, and this DL training in Tensorflow using
the Cifar100 dataset benefits from the GPU, the slope coefficient of the remote
predicted line is much smaller (4.85) than the slope of the predicted local line
(21.5), i.e., the local executions run 4.43x slower than the remote ones and,
after the intersection point ($e=7$) between the two predicted lines, the
migration pays off. This optimal epoch value is updated in the KB, which
concludes the Algorithm~\ref{alg:alg_parameters_update}. Finally, when the user
executes the actual cell containing the model fitting content, passing the
number epochs to train, the Migration Analyzer will use the recently updated
optimal epoch value for the migration decision.

\section{Related Work}

There are several efforts related to our work, in particular 
around Jupyter-based technologies, provenance of interactive notebooks, and
process migration. Here we highlight a few of them. Juric~\textit{et
al.}~\cite{juric2021checkpoint} present a solution for
checkpoint, restore, and live migration of Jupyter notebook sessions with
JupyterHub. Their motivation is to reduce the resource wastage from idle user sessions, while also exploring the use of cheaper resource types, such as spot instances, to reduce costs
hosting notebooks. With this capability, users can trigger complete session state checkpointing---which can be
restored on the same or an alternative resource.  Zonca and
Sinkovits~\cite{zonca2018deploying} describe strategies for deploying
Jupyter notebooks at the Extreme Science and Engineering Discovery Environment
(XSEDE). Their goal is to offer scalable hosting of Jupyter notebooks to
a large number of users via a distributed infrastructure. One of the deployment
strategies is to run Jupyter notebooks as jobs submitted to an HPC
scheduler. Users can then login into JupyterHub and directly reach a cluster node.
In this case, the job duration and other related
configuration options are provided by the system administrator or
set based on user preferences. Authors also explore Docker with
Swarm mode to explore the capability of having a web server
container running and not being affected by a node going down, as such
container can be migrated to another node transparently.

Pimentel~\textit{et al.}~\cite{pimentel2015collecting} introduce a provenance
system for interactive notebooks. Their motivation was that provenance support
in notebooks was limited to visualization of intermediate results and code
sharing, thus lacking the true history of all user interactivity. They
proposed a new provenance approach for notebooks based on noWorkflow, which is
a system to collect provenance from Python scripts. By doing so, they enable
users to reason about such provenance data and debug their work. The
noWorkflow system uses abstract syntax tree (AST) analysis to collect
definitions in notebooks so a provenance graph can be generated. The same
research team also performed a study on quality and reproducibility of Jupyter
notebooks~\cite{pimentel2019large}.
Also on provenance for Jupyter notebooks, Samuel and
K{\"o}nig-Ries~\cite{samuel2018provbook} introduce a tool called ProvBook,
which is a Jupyter extension to capture and view provenance data over the use
of the notebooks. With the tool, users can compare current and previous
experiments, querying sequences of executions using SPARQL\@.
From a training perspective, Smith~\textit{et al.}~\cite{smith2021towards}
introduce a study on modeling student engagement with Jupyter notebooks. For
their study, they consider metrics such as amount of time a student interacts
with the notebook, and the number of times students execute and modify cells.
Kery~\textit{et al.}~\cite{kery2018story} present a study on the usage of
Jupyter notebooks and how data scientists track their experiments.
Their focus was on the understanding of user behaviour, via a set of
Jupyter notebooks and how data scientists track of their experiments.
They focus on understanding user behaviour, via a set of
interviews, rather than a tooling to collect such data.
Process migration
across multiple computing platforms is a well studied topic in several
scenarios. For instance, there are efforts on seamlessly migrating virtual
machines on wide area networks~\cite{travostino2006livemigrationwan},
optimization of context migration from mobile device to a cloud
environment~\cite{li2016minimizing}, and offloading computation from
low-powered devices to remote resources accelerated by specialized hardware.
In this context, Zhang and Pande~\cite{zhang2005compiler} present a compiler
framework to minimize migration cost of application processes in a mobile
environment. They define a strategy that decides which parts of the program are
good migration points (so the program can continue executing elsewhere).  The
process relies on the instrumentation of the application (to keep track of file
operations, elapsed times, and memory usage) and on its profiling, from where
the reaching graph for the application is built. From that graph, they (1)
identify which files are accessed by which functions and (2) perform a backward
data flow analysis to track global variable liveness. With the information
learned about that program, the system computes the cost of migration at
potential candidate points and, as last step, the migration handler is inserted
in the code by the compiler so it can execute at run time.

Global variable liveness is also covered by Wang~\textit{et
al.}~\cite{wang2020bettercode}, which use ASTs to analyze the quality of
curated Python Jupyter Notebooks publicly available on a code hosting platform.
They identify unused variables and deprecated functions by extracting these
notebooks' ASTs and searching them for assignment operations (i.e.,
\emph{Store} nodes with names attached to them).  Names associated with
\emph{Store} operations but lacking an association with a \emph{Load} node are
considered unused objects and flagged as such. The authors conclude that poor
coding practices and lack of quality control is present in the majority of
those Notebooks. Different from the related work, we use ASTs to find data dependencies in
a Jupyter notebook's code cells for automatic migration between execution
environments, without requiring modifications of user code, while also
capturing provenance information, all initiated from a JupyterLab extension
that observes user interaction with notebooks.
\section{Conclusion}

We proposed a migration tool for Jupyter notebooks that migrates code to most
suitable computing platforms that increase user productivity.  The tool requires
no source code modification by the user, but only the installation of
a JupyterLab extension.  We also developed a component to reduce the notebook
state to optimize migration processes.
Our main findings developing the tool and performing experiments are that: (i)
Jupyter-related technologies enable a rich environment to add features to
increase user productivity; in our case, we enabled automatic execution of
notebook cells in remote platforms, annotating cells explaining migration
decisions; (ii) as notebooks are an extension of JSON, with JupyterLab allowing
us to extract information on when cells are executed and modified, it is
possible to build technology that explores the content of the cells and user
interactions to enable better migration decisions;
(iii) being an interactive platform, Jupyter notebooks cannot predict which
variables may be accessed in future cells, hence memory consumption tends to
grow over time; our AST-based notebook state reduction effectively works as
a context-aware garbage collector without compromising the creation of new
cells on that notebook;
(iv) to find the right
balance of computation power and monetary cost reduction, decisions on cell
migration should leverage understanding of user context, that is, common
sequences of cells normally executed by the user. This is because unnecessary
migrations can be avoided compared to migration decisions only based on
understanding computational needs of the current active notebook cell.

\bibliographystyle{IEEEtran}
\balance
\bibliography{references}

\end{document}